\documentclass[12pt,times,endnote]{article}
\usepackage{epsfig,float,color,graphicx}
\usepackage{amsmath}
\usepackage{amssymb}

\usepackage{amscd}
\begin{document}
\noindent
{\large\bf Effect of Fluctuations on the Freezing of a Colloidal Suspension 
in an External Periodic Potential}\\*[0.8cm]
\noindent
J. Chakrabarti($^{1,\star,\star\star}$\footnote{$^{(\star)}${~~~}Author for correspondence (e-mail: jaydeb\protect{@}amolf.nl)}$^{}$\footnote{$^{(\star\star)}${~~}{\it Present Address}: FOM Institute for Atomic and Molecular Physics, Kruislaan 407, 1098\\ \hspace*{1.3cm} SJ Amsterdam, The Netherlands.}) Supurna Sinha($^{2,\star\star\star}$\footnote{$^{(\star\star\star)}${~}{\it Present Address}: Raman Research Institute, Bangalore 560080, India})\\~\\

\noindent
($^{1}$) {\small Department of Physics, Indian Institute of Science, Bangalore 560 012, India}\\
\noindent
($^{2}$) {\small Jawaharlal Nehru Centre for Advanced Scientific Research and Centre for Theoretical Studies, Indian Institute of Science, Bangalore 560 012, India}\\*[0.2cm]

\noindent
{\small\it (Received 20 August 1996, received in final form 4 February 1997, accepted 7 February 1997)}.\\*[0.3cm]

\noindent
PACS.82.70.Dd -- Colloids\\
PACS.64.70.Dv -- Solid-liquid transitions\\*[0.3cm]

\noindent
{\small We incorporate the effects of fluctuations in a density functional analysis 
of the freezing of a colloidal liquid in the presence of an external potential 
generated by interfering laser beams. A mean -field treatment, using a density 
functional theory, predicts that with the increase in the strength of the 
modulating potential, the freezing transition changes from a first order to 
a continuous one via a tricritical point for a suitable choice of the modulating 
wavevectors. We demonstrate here that the continuous nature of the 
freezing 
transition at large values of the external potential $V_{e}$ survives the 
presence 
of fluctuations. 
We also show that fluctuations tend to stabilize the liquid 
phase in the large $V_{e}$ regime.} \\~\\*[0.8cm]

\section{Introduction}

\noindent
A few years ago Chowdhury, Ackerson and Clark [1] reported an interesting 
light scattering experiment on laser induced freezing of a two-dimensional 
suspension of strongly interacting colloidal particles. They showed that the 
colloidal liquid freezes into a two-dimensional crystalline phase with predominantly 
hexagonal order, when it is subject to a one-dimensional potential induced by a 
standing wave pattern of interfering laser beams. The wavevector $q_{o}$ 
of the modulating potential was chosen to coincide with the ordering wavevector 
of the colloidal liquid, {\it i.e.}, the location of the first peak of the 
direct correlation function (DCF) $c^{(2)}(q)$. This observation motivated 
experimental studies involving direct microscopic observations [2] and 
Monte-Carlo simulations [3] which confirmed the existence of such a laser 
induced freezing transition. However, the conclusions regarding the nature of 
the transition between the modulated liquid and the crystal have not been 
definitive. In the original paper where the experimental results were 
reported, Chowdhury {\it et al}.[1] also theoretically analyzed the 
phenomenon of laser induced freezing in terms of a simple phenomenological 
Landau-Alexander-McTague theory and concluded that the transition from the 
modulated liquid to the 2-d (modulated) crystalline phase can be made 
continuous for sufficiently large laser fields. The phenomenological Landau 
theory, based on a polynomial expansion of the free energy in powers of the 
order parameter truncated up to a finite degree, has well-known limitations 
in the case of a first order transition where the order parameter shows a 
jump. Apart from this, the coefficients in a phenomenological Landau type 
of free-energy are unknown and have unknown dependence on experimental 
parameters. In contrast the density functional theory (DFT) [4] pioneered by 
Ramakrishnan and Yusouff has led to a very successful approach for 
studying the strongly first order liquid-solid transitions for various 
systems including colloidal suspensions [5]. In contrast to the Landau-type 
free-energy functions, the density functional free-energy [4] 
is a non-polynomial function, (important) terms involving arbitrary powers 
in the order parameters being present. Furthermore, the coefficients of the 
density functional free energy are determined from the experimentally measured 
liquid correlation functions. It is of obvious interest to carry out 
DFT studies of Laser Induced Freezing. Recent density functional studies [6,7] 
have dealt with Laser Induced Freezing transition.\\

\noindent 
In reference [7] it has been shown how the freezing ({\it modulated liquid 
$\rightarrow$ crystal}) transition changes from a first order to a continuous 
one via a tricritical point with the increase in the external potential $V_{e}$, 
if the modulation wave-vector is tuned at the ordering wavevector of the liquid. 
The order parameters in this theory [7] are the Fourier components of the 
molecular field $\xi ({\bf r})(\equiv \ln (\rho({\bf r})/\rho_{0}))$ with 
wavevectors equal to the reciprocal lattice vectors (RLV) of the periodic 
structure, into which the liquid would have frozen in the absence of $V_{e}$. 
Here $\rho(\rm r)$ is the local density in the modulated liquid or the crystal and 
$\rho_{0}$ is the mean density of the liquid. Symmetry considerations show that in 
the presence of $V_{e}$, the order parameters corresponding to the smallest set of 
RLVs of the crystal divide into two classes: (1) those corresponding to the 
RLVs parallel to the modulation wave-vectors $\{{\bf g}_{f}\}$ of $V_{e}$ denoted 
by $\xi_{f}$; and (2) those corresponding to the rest of the RLVs in the 
smallest set $\{{\bf g}_{d}\}$ denoted by $\xi_{d}$. One can choose the wavevectors 
pertaining to the external potential in such a way that an integral combination 
of vectors in the class $\{{\bf g}_{f}\}$ cannot be obtained from an odd combination 
of vectors in the class $\{{\bf g}_{d}\}$. Under this condition it is shown [7] that 
the Landau free energy obtained by expanding the DFT free energy about the 
modulated liquid phase ($\xi_{f} \neq 0$) in powers of $\xi_{d}$ has only even 
powers of $\xi_{d}$ and the transition changes from first order for low values 
of $- \beta V_{e}$ [7] (where the quartic coefficient of the Landau free energy 
is negative) to a continuous one for large values of $- \beta V_{e}$ [7] (where 
the quartic coefficient is positive). Note that this symmetry argument 
holds good under the inclusion of arbitrarily large RLV and liquid direct 
correlation functions of any order [7].\\

\noindent
The analysis of reference [7], however, is of a mean field nature. We clarify 
below in what precise sense the theory presented in reference [7] is a mean 
field theory. The density $\rho({\bf r})$, or equivalently the molecular 
field $\xi({\bf r})$, contains the freezing modes, {\it i.e}., the modes with 
wavevectors equal to the reciprocal lattice vectors (RLV) of the periodic 
structure (into which the liquid would have frozen in the absence of 
$V_{e}$), and other modes which we call the fluctuating modes. So we can write 
that $\xi({\bf r}) = 
\sum_{\bf g}\xi_{\bf g} \exp(i{\bf g} \cdot {\bf r})
+\sum_{\bf q}^{/}\xi_{\bf q} \exp(i{\bf q} \cdot {\bf r})$ 
where the first summation runs over the freezing modes and the second summation 
runs over the fluctuating modes. In the present problem, for instance, $\xi_{f}$ 
and $\xi_{d}$ are the freezing modes corresponding to the smallest RLVs 
parallel to the wave-vectors of $V_{e}$ and the rest of the RLVs in the 
smallest set respectively. In reference [7] only the freezing modes are 
retained and all the fluctuating modes are completely neglected. Thus the 
theoretical analysis presented in reference [7] is a mean field one. 
In general, the fluctuating modes can have an important effect on the 
freezing transition. This motivates us to extend the mean field treatment 
of reference [7] by including the fluctuating modes in addition to the 
freezing modes which are already present in the mean field study of 
reference [7]. In this context we recall the work of Brazovskii [9-11].\\
\noindent
He considered the transition from an isotropic liquid to a cholesteric liquid 
crystal. Here the mean field Landau free energy in terms of the order parameter 
contains only even powers and $T^{(4)}$, the coefficient of the quartic 
term is 
positive; hence the transition is predicted to be continuous. He demonstrated 
that fluctuations transform a continuous transition to a first order one. 
The large effect of fluctuations in Brazovskii's theory is due to a considerable 
softening of the order parameter modes at a nonzero value $q_{0}$ of the 
wavevector. As a result, the effect of fluctuations comes from a large surface 
area $(\simeq 4\pi q_{0}^{2})$ in the momentum space. After integrating 
out 
these fluctuation modes, the coefficient of the quartic term in the effective 
Landau free energy becomes negative where the coefficient of the quadratic 
term changes sign, indicating that the continuous freezing transition is 
preempted by a first-order one. So an obvious point to investigate is to what 
extent the fluctuating wavevectors close to the ordering wavevector $q_{0}$ 
of the liquid are important in the presence of an external modulation 
$- \beta V_{e}$ tuned at $q_{0}$ [1-3,7]. More specifically we ask if the quartic 
coefficient in the effective Landau expansion can become negative (where the 
quadratic coefficient changes sign) in the context of Laser Induced Freezing 
transition in the large $- \beta V_{e}$ limit, where the transition is predicted 
to be continuous in reference [7]. We consider a two-dimensional colloidal 
system subject to a one-dimensional external potential for the sake of 
simplicity. Furthermore, it is well-known that the effect of fluctuations 
is more prominent in two dimensions than in three dimensions. \\

\noindent
Our starting point is the free-energy cost of creating a density inhomogeneity 
over a uniform liquid (in the incompressible limit), in terms of $\xi_{\bf 
q}$ [7,8]: 

\begin{eqnarray}
\beta F(\{\xi_{\bf q}\}) = &-& \ln \left[ \frac{1}{V} \int {\rm d}^{d}r 
\exp[\xi_{\rm f} \sum_{J} e^{(i{\bf g}_{\rm f}^{(J)} {\bf r})}]
\exp[\sum_{\bf q} \xi_{\bf q}e^{i{\bf q}\cdot{\bf r}}]\right]\nonumber \\
&+&\frac{1}{2}\sum_{q} \xi^{2}_{q}/c^{(2)}_{\bf q}+n_{\rm f}\beta V_{e}\xi_{\rm f}/c^{(2)}_{1}
\end{eqnarray}
where the summation over {\bf q} includes the set $(\{{\bf g}_{\rm d}\})$ 
and the 
fluctuating modes. Here $c^{(2)}_{\bf q}$ is the direct correlation function of 
the liquid, $V$ is the volume of the system and $n_f$ is the number of 
vectors in 
the class $(\{{\bf g}_{\rm f}\})$. We generate a Landau expansion from 
equation 
(1) about the modulated liquid phase $(\xi_{\rm f} \neq 0)$ in powers of $\xi_{\bf q}$
(where the transition is continuous according to the mean field theory) and then 
integrate out the fluctuating modes to form an effective Landau expansion in powers of 
$\xi_{\rm d}$. This free energy has been used to obtain the phase diagram. 
Note that the odd invariants of $\xi_{\rm d}$ are forbidden in the effective 
Landau free-energy due to symmetry restrictions as in the mean field case [7]. 
In this work we follow Brazovskii's approach to form the effective Landau 
free-energy. In particular, Brazovskii shows that the most dominant corrections to 
the two-point and four-point correlation functions come from diagrams with a 
single loop. As a consequence, it is sufficient to evaluate self consistently 
the self-energy corrections to one-loop order. Similarly, it is shown [9] 
that the corrections to the four-point vertex functions are dominated by a 
restricted class of one-loop diagrams and their ladders. Here we have a liquid 
of strongly interacting colloidal particles characterized by a static structure 
factor $S(q)$, with a sharp maximum at $q\simeq q_{0}$. Equivalently, this 
implies that $[S(q)]^{-1}$ has a prominent minimum at $q\simeq q_{0}$, a 
feature 
present in Brazovskii's theory as well. However, in this situation there are a 
few additional complications which are absent in Brazovskii's problem. Here we 
consider a phase transition induced by an external potential. The coupling of 
the external potential to certain density modes of the isotropic liquid leads 
to a nontrivial structure of the correlation matrix in the modulated liquid 
phase which plays a central role in the computation of the renormalized 
four-point vertex functions. Furthermore, the momentum conservation laws 
appearing in the present work involve the additional wavevectors pertaining 
to the external modulation. Also, the bare vertex functions appearing in our 
free energy have a complicated dependence on the external potential. For the 
sake of simplicity we confine ourselves to asymptotically large values of the 
external potential. In this limit, we neglect fluctuations in the modes $\xi_{\rm f}$ 
(pertaining to the external potential) since a zero temperature 
normal mode analysis shows that the external potential creates a gap 
$(\Delta \simeq |V_{E}|)$ in the phonon spectrum along the direction of 
externally induced ordering. Consequently fluctuations in these order parameter 
modes are energetically unfavorable for $|\beta V_{e}| \gg 1$. We therefore 
treat the modes $\xi_{\rm f}$  non-perturbatively [7]. We calculate the correlation 
matrix in the modulated liquid phase by considering one-loop Hartree corrections 
to the self-energy (Sect. 3). It is easy to evaluate the coefficient ($T_{2}$) of 
the quadratic term in this effective Landau free energy from a knowledge of the 
correlation matrix in the modulated liquid phase (Sect. 3). We determine the 
point in the phase diagram where $T_{2}$ goes to zero. We then use the correlation 
matrix to evaluate the renormalized four-point vertex function and consequently 
$T_{4}$, the coefficient of the quartic term in the effective Landau free energy 
at that point (Sect. 4). We find that $T_{4}$ remains positive in our regime of 
interest indicating that the freezing transition remains continuous for large 
values of $V_{e}$ even in the presence of fluctuations. We also show, as we will 
elaborate later, that fluctuations tend to stabilize the liquid phase relative 
to the solid phase in the limit of large $V_{e}$.\\

\noindent 
The paper is organized as follows. In Sections 2, 3 and 4, we give a detailed 
account of our calculation leading to the determination of the nature of the 
laser induced freezing transition and in Section 5 we present our numerical results. 
We conclude the paper in Section 6 with a few remarks. \\

\section{Free Energy}

\noindent
Expanding the free-energy in equation (1), in a power series and 
retaining terms to quartic order in $\xi_{\bf q} ({\bf q} \neq {\bf 
g}_{\rm f}$ 
or zero), we get:
\begin{eqnarray}
\beta F(\{\xi_{\bf q}\}) &=& \frac{1}{2} \sum_{\bf q} \xi^{2}_{\bf q}/c^{(2)}_{\bf q} - 
\sum_{{\bf q}_{1},{\bf q}_{2}} \frac{1}{2} A_{{\bf q}_{1},{\bf q}_{2}}\xi_{{\bf q}_{1}}\xi_{{\bf q}_{2}} - \frac{1}{6}\sum_{{\bf q}_{1},{\bf q}_{2},{\bf q}_{3}} T_{{\bf q}_{1},{\bf q}_{2},{\bf q}_{3}}\xi_{{\bf q}_{1}}\xi_{{\bf q}_{2}}\xi_{{\bf q}_{3}}\nonumber \\
&-&\frac{1}{24} \sum_{{\bf q}_{1},{\bf q}_{2},{\bf q}_{3},{\bf q}_{4}} Q_{{\bf q}_{1},{\bf q}_{2},{\bf q}_{3},{\bf q}_{4}} \xi_{{\bf q}_{1}}\xi_{{\bf q}_{2}}\xi_{{\bf q}_{3}}\xi_{{\bf q}_{4}}
\end{eqnarray}
where
\begin{eqnarray}
A_{{\bf q}_{1},{\bf q}_{2}} &=& \frac{\int{\rm d}^{d}r
\exp[\xi_{\rm f} \sum_{j}e^{(i{\bf g}_{\rm f}^{(j)})\cdot {\bf 
r}}]e^{{i}(\sum_{i=1\;\;2}{\bf q}_{i})\cdot {\bf r}}}
{\int{\rm d}^{d}r\exp[\xi_{\rm f} \sum_{j}e^{(i{\bf g}_{\rm f}^{(j)} {\bf r})}]}\nonumber\\
T_{{\bf q}_{1},{\bf q}_{2},{\bf q}_{3}} &=& \frac{\int{\rm d}^{d}r
\exp[\xi_{\rm f} \sum_{j}e^{(i{\bf g}_{\rm f}^{(j)} {\bf 
r})}]e^{{i}(\sum_{i=1\;\;3}{\bf q}_{i})\cdot {\bf r}}}
{\int{\rm d}^{d}r\exp[\xi_{\rm f} \sum_{j}e^{(i{\bf g}_{\rm f}^{(j)} {\bf r})}]}\nonumber\\
Q_{{\bf q}_{1},{\bf q}_{2},{\bf q}_{3},{\bf q}_{4}} &=& Q^{(1)}_{{\bf q}_{1},{\bf q}_{2},{\bf q}_{3},{\bf q}_{4}} + Q^{(2)}_{{\bf q}_{1},{\bf q}_{2},{\bf q}_{3},{\bf q}_{4}}
\end{eqnarray}
with
\begin{eqnarray}
Q^{(1)}_{{\bf q}_{1},{\bf q}_{2},{\bf q}_{3},{\bf q}_{4}} &=& 
\frac{\int{\rm d}^{d}r[\xi_{\rm f} \sum_{j}e^{(i{\bf g}_{\rm f}^{(j)} {\bf r})}
]e^{{i}(\sum_{i=1,4}{\bf q}_{i})\cdot {\bf r}}}
{\int{\rm d}^{d}r\exp[\xi_{\rm f} \sum_{j}e^{(i{\bf g}_{\rm f}^{(j)} {\bf r})}]}\nonumber\\
Q^{(2)}_{{\bf q}_{1},{\bf q}_{2},{\bf q}_{3},{\bf q}_{4}} &=& -3A_{{\bf q}_{1},{\bf q}_{2}} A_{{\bf q}_{3},{\bf q}_{4}}.
\end{eqnarray}

\noindent
One can easily verify that each of these coefficients is nonzero if and only if the 
sum of the {\bf q}-vectors appearing in the coefficient is {\bf G}$_{\rm f}$, 
which is an arbitrary integral combination of vectors $\{{\bf g}_{\rm f}\}$. This is 
the momentum conservation condition for the present problem. For instance, 
$A_{{\bf q}_{1},{\bf q}_{2}}$ in equation (3) will be nonzero if 
${\bf q}_{1} + {\bf q}_{2} = {\bf G}_{\rm f}$. The cubic term in equation (3) will 
have nonzero value if $\sum_{i=1,3} {\bf q}_{i} = {\bf G}_{\rm f}$. Similarly 
$Q^{(1)}_{{\bf q}_{1{\bf q}_{2},{\bf q}_{3},{\bf q}_{4}}}$ in equation (4) 
is nonzero if $\sum_{i=1,4} {\bf q}_{i} = {\bf G}_{\rm f}$.\\

\noindent
The mean field calculation shows that in the absence of $V_{e}$ a 
2-dimensional liquid 
freezes into a triangular lattice [1, 12] with its 6 smallest reciprocal lattice 
vectors (RLV) $\{{\bf g}_{0}\} = (0, \pm 1)q_{0}, (\pm 1/2, \pm \sqrt{3}/2)q_{0}$.
Here $q_{0}$, as is known from DFT, corresponds to the first peak of DCF. We choose 
the wavevectors of $V_{e}$ to be $\{{\bf g}_{\rm f}\} = (0, \pm 1)q_{0}$ in this set. 
This choice simplifies the geometry of the problem greatly. 
With this choice, the 
vector ${\bf G}_{\rm f}$ can be written as $m{\bf g}_f$ where $m$ is an 
arbitrary integer and 
${\bf g}_{\rm f} = (0,1)q_{0}$. Since there is a one-dimensional ordering created 
by $V_{e}$, there is a periodicity in the x-direction. This symmetry of the 
problem enables us to write a general wavevector {\bf p} as a sum of {\bf q}, a vector 
in the first Brillouin Zone (BZ) and an arbitrary number of ${\bf g}_{\rm f}$.  
Here the first BZ is the strip between $q_{y} = -0.5$ to $q_{y} = 0.5$. Now we 
introduce some simplifications in our notation. Let us consider, for instance, 
$A_{{\bf q}_{1},{\bf q}_{2}}$ defined in equation (3). Note that the momentum 
conservation demands that ${\bf q}_{1} + {\bf q}_{2} = m{\bf g}_{\rm f}$. 
So if we choose a vector {\bf q} in the first Brillouin Zone such that 
${\bf q}_{1} = {\bf q} + i{\bf g}_{\rm f}$, then the momentum conservation 
demands that ${\bf q}_{2} = -{\bf q} + j{\bf g}_{\rm f}$, so that $i + j = m$. 
So we define 
$A_{{\bf q}_{1}, {\bf q}_{2}} \equiv A_{({\bf q} + i{\bf g}_{\rm f}),(-{\bf q}+j{\bf g}_{\rm f})} \equiv A_{ij}({\bf q})$, where the momentum conservation law 
$\sum_{i} {\bf q}_{i} = m{\bf g}_{\rm f}$, is automatically satisfied. 
One can write similar simplified notations for the cubic and the quartic 
vertices.\\

\noindent
In the large $-\beta V_{e}$ limit, the integrals appearing in the various 
coefficients 
in equation (3) can be evaluated by means of an asymptotic expansion. Let us consider 
the vertex in equation (3). For the 2-d hcp lattice and the given choice of 
${\bf g}_{\rm f}$, we have $\phi_{\rm f} \equiv \sum_{i} \exp (i{\bf g}_{\rm f}^{(i)}\; {\bf r}) = 2\cos (\frac{4\pi}{\sqrt{3}} x)$.  
In the large $-\beta V_{e}$ limit one can expand 
$\phi_{\rm f}$ about its maximum and the resulting Gaussian integral can be 
computed exactly. In this way the following expressions can be derived:
\begin{eqnarray}
A_{ij} &=& \exp \left[- \frac{3(i+j)^{2}}{64\pi^{2}\xi_{\rm f}}\right] \nonumber\\
T_{ijk} &=& \exp \left[- \frac{3(i+j+k)^{2}}{64\pi^{2}\xi_{\rm f}}\right] \nonumber\\
Q_{ijkl}^{(1)} &=& \exp \left[- \frac{3(i+j+k+l)^{2}}{64\pi^{2}\xi_{\rm f}}\right] \nonumber\\
Q_{ijkl}^{(2)} &=& 3\exp \left[- \frac{3(i+j)^{2}}{64\pi^{2}\xi_{\rm f}}\right]
\exp\left[- \frac{3(k+l)^{2}}{64\pi^{2}\xi_{\rm f}}\right].
\end{eqnarray}

\noindent
In the large $-\beta V_{e}$ limit the vertices do not depend on the momentum labels. 
Note that $-\beta V_{e}$ enters the calculation through $\xi_{\rm f}$. In equation (1) 
we set $\xi_{\bf q} = 0$ for all {\bf q} except ${\bf q} = {\bf g}_{\rm f}$ and make an
asymptotic expansion as above. The self-consistency condition at the minimum of
the resulting free energy with respect to $\xi_{\rm f}$ yields the following expression 
for $\xi_{\rm f}$:
\begin{eqnarray*}
\xi_{\rm f} = -\beta V_{e} + \rho_{0}c^{(2)}(q_{0}) + 1/(4\beta V_{e})
\end{eqnarray*}
to ${\cal O}(1/\beta V_{e})$ in this limit.\\

\section{Correlation Matrix}

\noindent
The first step is to calculate the bare correlation matrix for the modulated liquid 
phase. Since the effect of the external potential is to scatter a mode with momentum 
{\bf q} into that with 
\begin{center}
\includegraphics*[width=3.8in]{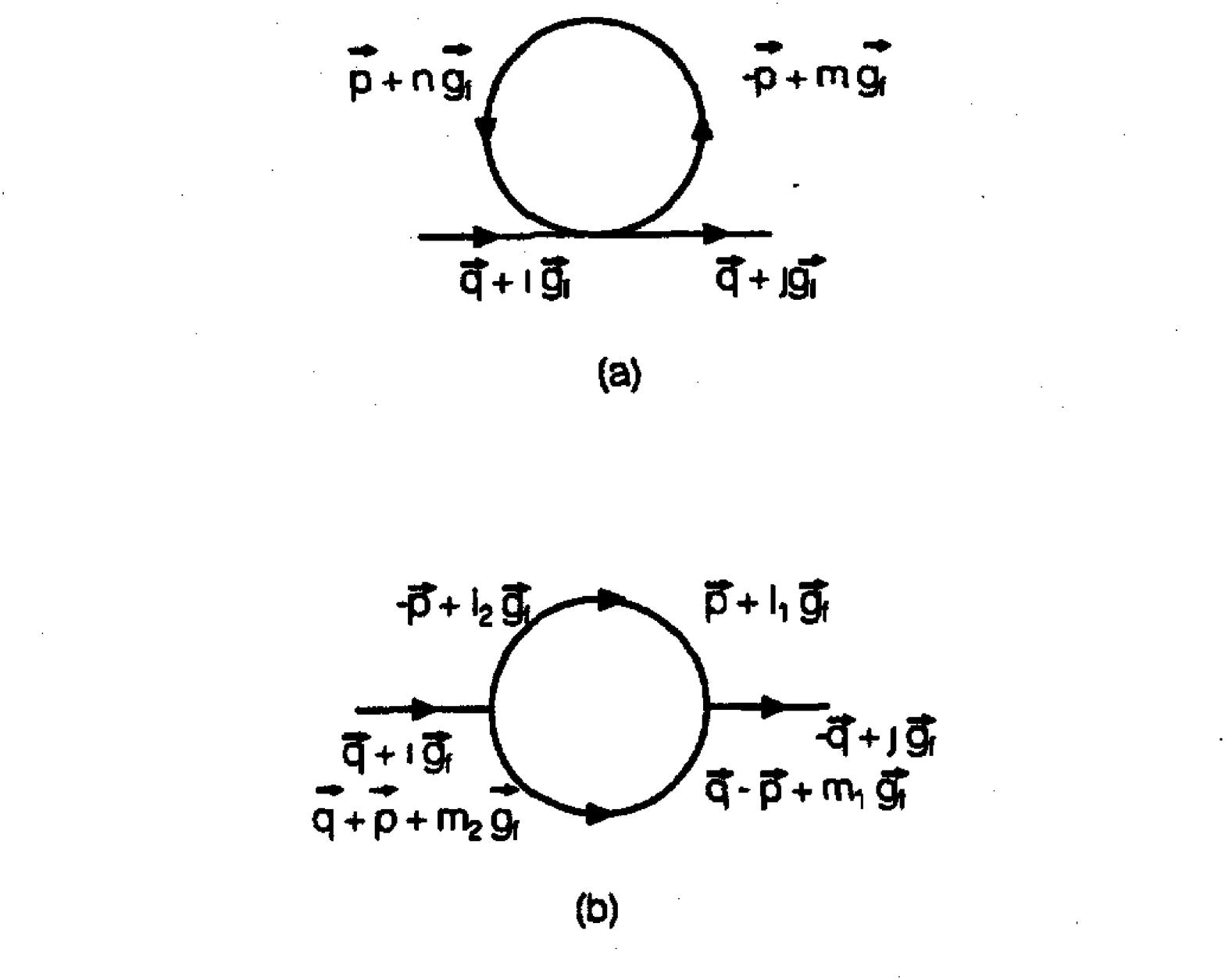}
\end{center}
\vspace*{0.3cm}
Fig. 1.~~Lowest order correction to the self-energy due to (a) the 4-point vertices 
and (b) the 3-point vertices.\\*[0.2cm]

\noindent
momentum ${\bf q} + m{\bf g}_{\rm f}$, where $m$ is any integer, the quadratic 
part of the free-energy equation (2) has off-diagonal couplings, the strength of 
which is given by $A_{ij}$ in equation (5). The elements of this matrix are: 
\begin{eqnarray}
\frac{1}{\rho_{0}c^{(2)} (|{\bf q} + i{\bf g}_{\rm f}|)} - 1 
&& {\rm (diagonal)}\nonumber\\
-\exp \left( - \frac{3(i+j)^{2}}{64\pi^{2}\xi_{\rm f}}\right) && {\rm (off 
- diagonal)}
\end{eqnarray}

\noindent
The bare correlation matrix ${\cal G}_{ij}^{(0)}$ for the modulated phase 
is 
obtained in terms of 
$c^{(2)}(q)$ and $-\beta V_{e}$ (via $\xi_{\rm f}$) by numerically 
inverting the matrix 
in equation (6). The main difficulty is that the correlation matrix is of infinite order.\\

\noindent 
At this point we introduce some approximations that enable us to deal with a 
sufficiently simple and finite dimensional correlation matrix. We estimate the 
contributions of the corrections coming from the cubic and the quartic vertices. 
To the lowest order the correction due to the quartic vertex is given by: 
\begin{eqnarray}
\sum^{\rm H}_{ij} = \frac{1}{2(2\pi)^{2}} \sum_{m,n} Q_{ijmn} \int {\rm d}{\bf p} {\cal G}^{(0)}_{mn} ({\bf p}).
\end{eqnarray}

\noindent
This correction does not depend on the external momentum label {\bf q} (see Fig. 1a). 
Similarly, the lowest order contribution due to the cubic vertex is:
\begin{eqnarray}
\sum^{\rm F}_{ij} ({\bf q}) = \frac{1}{2(2\pi)^{2}} \sum_{l_{1},l_{2},m_{1},m_{2}} 
T_{il_{2}m_{2}} T_{jl_{1}m_{1}} \int {\rm d}{\bf p} {\cal G}^{(0)}_{l_{1}l_{2}}
({\bf p}) {\cal G}^{(0)}_{m_{1}m_{2}} ({\bf q} - {\bf p}).
\end{eqnarray}

\noindent
This correction, however, depends on the external momentum {\bf q} (see Fig. 1b). 
The integrations in both cases are over the first BZ. We determine the more 
dominant of these two self-energy corrections. Near $q_0$ we take a 
simpler form 
for the liquid direct correlation function: 
\begin{eqnarray}
[1/\rho_{0}c^{(2)} (|{\bf q}|)] - 1 = c_{0} + d(q-q_{0})^{2},
\end{eqnarray}
where $c_{0}$ and $d$ parametrize the liquid direct correlation function. \\
\noindent
Note that:
\begin{eqnarray*}
\rho_{0}c^{(2)}(q_{0}) = 1/(1 + c_{0}),
\end{eqnarray*} 

\noindent
{\it i.e.}, $c_{0}$ pertains to the first peak height of the liquid direct 
correlation function and $d$ corresponds to the width]. We compute 
$\sum^{\rm H}_{ij}$ (Eq. (7)) and (Eq. (8)), using the bare correlation 
matrix (see Eq. (6)). We go to the diagonal basis and compare the eigenvalues 
of $\sum_{\rm H}^{ij}$ and $\sum^{\rm F}_{ij}$ for various {\bf q}. It turns 
out that the most dominant eigenvalue of $\sum^{\rm f}_{ij} ({\bf q})$ 
gives smaller corrections than that of $\sum^{\rm H}_{ij}$ for all 
$|{\bf q}|$ lying in the region near $q_{0}$. This means that the 
predominant correction comes from $\sum^{\rm H}_{ij}$ for $|{\bf q}| \sim q_{0}$. 
We also observe that the most dominant eigenvalue does not change 
significantly with increased dimensionality of $\sum^{\rm H}_{ij}$ beyond 
$3 (i = \pm 1$ and $j = \pm 1)$. To summarize, we confine our calculations 
to the region $|{\bf q}| \sim q_{0}$ ({\it i.e.}, we confine to fluctuations 
in the low energy modes of the order parameter spectrum), keeping only 
$\sum^{\rm H}_{ij}$ in the self-energy correction and restrict to the block 
$(i,j) \in (-1,1)$ of $\sum^{\rm H}_{ij}$. \\

\noindent
Notice that the off-diagonal elements of ${\cal G}^{-1}_{0}$ depends only 
on the magnitude of $(i,j)$ and not on $i$ and $j$ separately. This means 
that there are only two independent off-diagonal elements. Hence within 
our approximation the $3 \times 3$ bare correlation matrix is as displayed 
below: 
\[
{\cal G}^{-1}_{0} = \left[ \begin{array}{ccc}
c_{0} + d(|q - g_{\rm f}| - q_{0})^{2} & a_{0} & b_{0}\\
a_{0} & c_{0} + d(q - q_{0})^{2} & a_{0}\\
b_{0} & a_{0} & c + d(|q + g_{\rm f}| - q_{0})^{2}\end{array}\right].
\]
Here $a_{0} = -\exp[-3/64\pi^{2}\xi_{\rm f}]$ is the off-diagonal coupling 
with $(i+j) = 1$. Similarly, $b_{0} = -\exp[-(12/64\pi^{2}\xi_{\rm f})]$ 
is the off-diagonal coupling with $(i+j) = 2$. We refer to $c_{0}$, $a_{0}$, 
$b_{0}$ and $d$ as the bare parameters.\\

\noindent 
Dyson's equation for the corrected correlation matrix ${\cal G}_{ij}$ is: 
\begin{eqnarray}
[{\cal G}_{ij} ({\bf q})]^{-1} = [{\cal G}^{(0)}_{ij} ({\bf q})]^{-1} - \sum_{ij} ({\bf q})
\end{eqnarray} 
where 
\begin{eqnarray}
\sum_{ij} ({\bf q}) \equiv \sum^{\rm H}_{ij} = \frac{1}{2(2\pi)^{2}} \sum_{mn} Q_{ijmn} \int {\rm d}{\bf p}{\cal G}_{mn}({\bf p}).
\end{eqnarray}
Since the correction $\sum^{\rm H}_{ij}$ does not depend on {\bf q}, the {\bf q}-dependent 
term in the bare and the corrected correlation matrix are the same ({\it i.e.}, $d$ is 
unaffected by the correction). The bare parameter $c_{0}$ gets renormalized to $c$ as a 
result of the correction and so do the off-diagonal elements. Hence, 
${\cal G}^{-1}$ is a $3\times 3$ matrix:
\[
{\cal G}^{-1} = \left[ \begin{array}{ccc}
c + d(|q - g{\rm f}| - q_{0})^{2} & a & b\\
a & c + d(q - q_{0})^{2} & a\\
b & a & c + d(|q + g_{\rm f}| - q_{0})^{2}\end{array}\right],
\]
where now $a$ and $b$ are renormalized off-diagonal couplings. Hence the self-consistency 
equations derived from equation (10) involve three nontrivial parameters $c$, $a$ and $b$. 
It is clear that in order to frame these equations it is sufficient to consider 
diagrams pertaining to momentum transfers ({\it i.e}. $(i+j)$) of magnitude $0$, $1$ and $2$ 
only. These contributions are denoted by $\sum_{(0,0)}$ (See Fig. 2), 
$\sum_{(-1,0)}$ and $\sum_{(-1,1)}$ respectively. The self-consistency equations 
\begin{eqnarray}
c &=& c_{0} - \sum_{(0,0)},\nonumber\\
a &=& a_{0} - \sum_{(-1,0)},\nonumber\\
b &=& b_{0} - \sum_{(-1,1)},
\end{eqnarray}
\vspace*{0.5cm}
\begin{center}
\includegraphics*[width=3.8in]{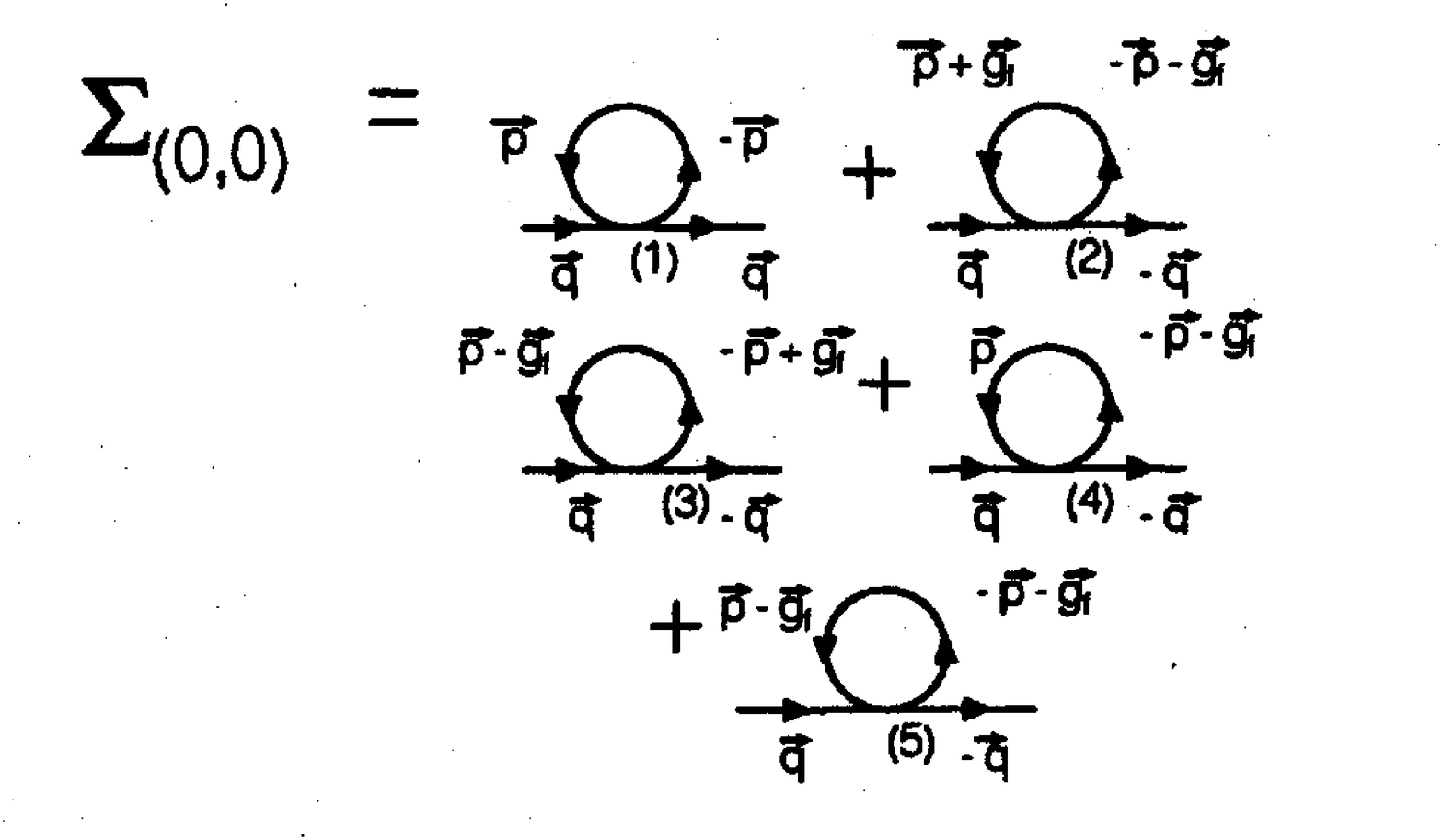}
\end{center}
\vspace*{0.3cm}

\noindent
Fig. 2. - The diagrams which contribute to $\sum_{(0,0)}$. \\*[0.3cm]

\noindent
where $\sum_{(0,0)}, \sum_{(-1,0)} {\rm and} \sum_{(-1,1)}$ are given by:
\begin{eqnarray}
\sum_{(0,0)} &=& \int {\rm d}{\bf p}[R({\cal G}_{0,0}({\bf p})
+ {\cal G}_{1,1}({\bf p}) + {\cal G}_{-1,-1}({\bf p})) + {\cal SG}_{1,0}({\bf p}) + {\cal TG}_{-1,1}({\bf p})]\nonumber\\
\sum_{(-1,0)} &=& \int {\rm d}{\bf p}[U({\cal G}_{0,0}({\bf p}) 
+ {\cal G}_{1,1}({\bf p})+ {\cal G}_{-1,-1}({\bf p})) + V{\cal G}_{1,0}({\bf p}) + {\cal WG}_{-1,1}({\bf p})]\nonumber\\
\sum_{(-1,1)} &=& \int {\rm d}{\bf p}[M({\cal G}_{0,0}({\bf p}) 
+ {\cal G}_{1,1}({\bf p})
+ {\cal G}_{-1,-1}({\bf p}) + {\cal NG}_{1,0}({\bf p}) + {\cal PG}_{-1,1}({\bf p})].
\end{eqnarray}
\noindent
Here,

\begin{eqnarray}
R &=& 1/(2\pi)^{2}\nonumber\\
S &=& 4\exp(-3/64\pi^{2}\xi_{\rm f})/(2\pi)^{2}\nonumber\\
T &=& 2\exp(-12/64\pi^{2}\xi_{\rm f})/(2\pi)^{2}\nonumber\\
U &=& \exp(-3/64\pi^{2}\xi_{\rm f})/(2\pi)^{2}\nonumber\\
V &=& (-2\exp(-12/64\pi^{2}\xi_{\rm f}) + 12\exp(-6/64\pi^{2}\xi_{\rm f}) + 3\exp(-1/64\pi^{2}\xi_{\rm f}) - 2)/(2\pi)^{2}\nonumber\\
W &=& (-\exp(-27/64\pi^{2}\xi_{\rm f}) + 6\exp(-15/64\pi^{2}\xi_{\rm f}) -\exp(-3/64\pi^{2}\xi_{\rm f}))/2(2\pi)^{2}\nonumber\\
M &=& \exp(-12/64\pi^{2}\xi_{\rm f})/(2\pi)^{2}\nonumber\\
N &=& (-2\exp(-3/64\pi^{2}\xi_{\rm f}) + 12\exp(-15/64\pi^{2}\xi_{\rm f}) - 2\exp(-27/64\pi^{2}\xi_{\rm f}))/2(2\pi)^{2}\nonumber\\
P &=& (-4\exp(-24/64\pi^{2}\xi_{\rm f}) - \exp(-48/64\pi^{2}\xi_{\rm f}) - 1)/2(2\pi)^{2}
\end{eqnarray}

\noindent
Let us consider diagram (4) in Figure 2 for the purpose of illustration. The sum 
of the momenta at the vertex is $-{\bf g}_{\rm f}$. Hence the contribution of 
$Q^{(1)}$ is $\-exp(-3/64\pi^{2}\xi_{\rm f})$. To calculate the contribution from 
$Q^{(2)}$, we note that the sum of the momenta of the two free legs is zero and 
that of the internal lines is $-{\bf g}_{\rm f}$. So the contribution is 
$-3\exp(-3/64\pi^{2}\xi_{\rm f})$. Adding the two contributions and multiplying the 
sum by the symmetry factor and $\frac{1}{2\pi^{2}}$ coming from the phase volume, 
one gets the expression for $S$ in $\sum_{(0,0)}$. Similar consideration yields 
the other numerical coefficients. \\

\noindent
We solve the coupled equations in equation (12) numerically for $c$, $a$ and $b$ 
for a given $c_{0}$ and $-\beta V_{e}$ ({\it i.e.}, given bare parameters $c_{0}$, 
$a_{0}$ and $b_{0}$). From the self-consistent values of $c$, $a$ and $b$, one 
has to construct the coefficient of the term $(\xi_{\rm d})^{2}$ in the effective 
Landau free-energy. To this end we enumerate the freezing modes as follows: 
${\bf g}_{\rm d}^{(1)} = (\sqrt{3}/2) (1,1/\sqrt{3})$, ${\bf g}_{\rm d}^{(2)} = (\sqrt{3}/2)(1,-1/\sqrt{3})$, ${\bf g}_{\rm d}^{(3)} = (\sqrt{3}/2)(-1,-1/\sqrt{3})$,

\begin{center}
\includegraphics*[width=3.8in]{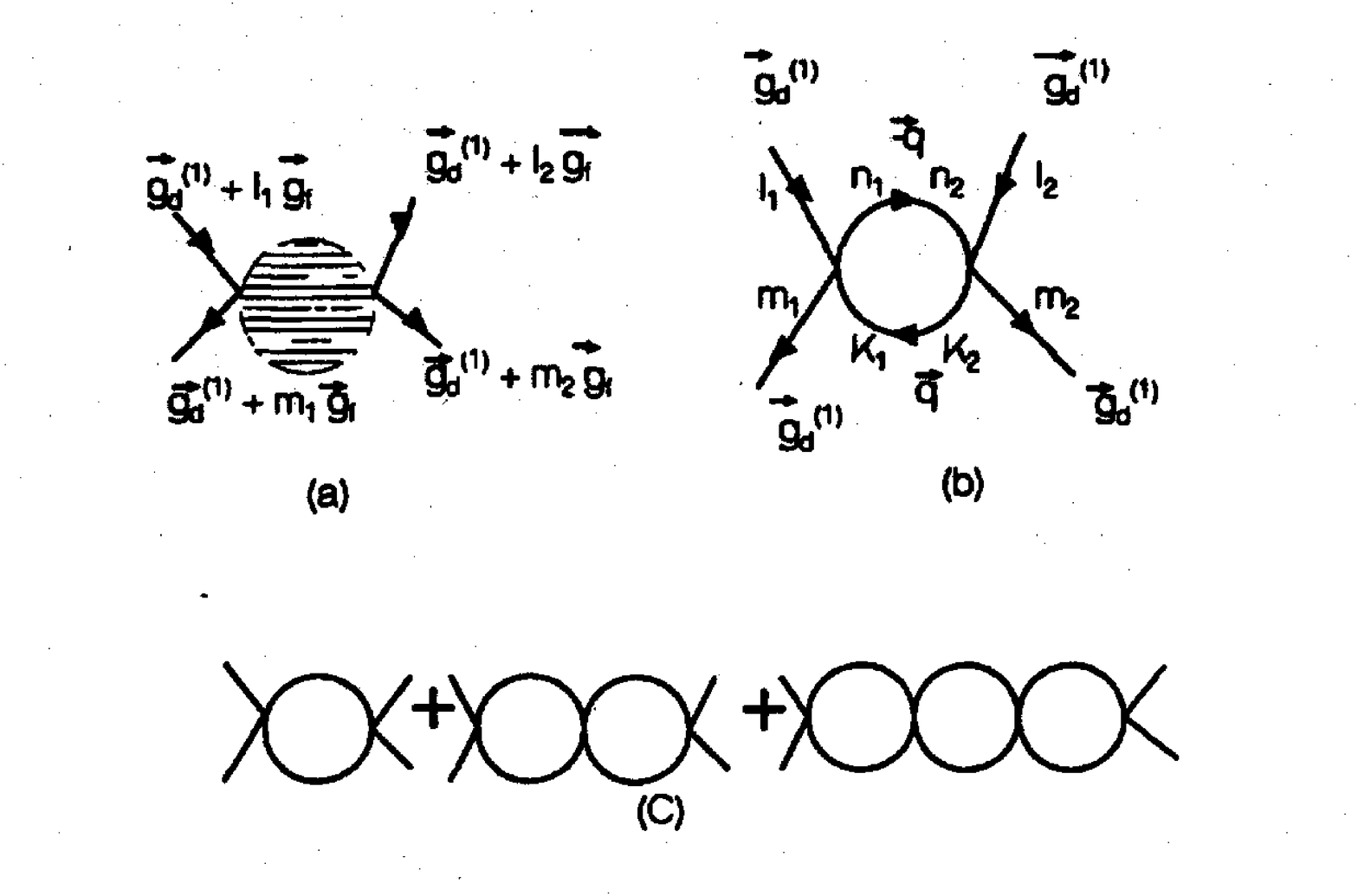}
\end{center}
\vspace*{0.3cm}
\noindent
Fig. 3. - (a) The choice of the external legs for a typical 4-point vertex 
that appears in the quartic term in the effective Landau free energy. 
(b) A one-loop correction to a 4-point vertex with external legs as in (a). 
The labels on the lines indicate the number of ${\bf g}_{\rm f}$ to be 
added to the momentum concerned, for instance label $l_{1}$ indicates that 
the momentum is ${\bf g}^{(1)}_{\rm d} + l_{1}{\bf g}_{\rm f}$. 
Similarly the labels $n_{1}$ and $n_{2}$ on the internal line indicates 
the $n_{1}n_{2}{\rm th}$ element of the ${\cal G}$ matrix. (c) The ladder 
corresponding to (b). \\*[0.3cm]

\noindent
${\bf g}_{\rm d}^{(4)} = (\sqrt{3}/2)(-1,1/\sqrt{3})$. If we assume that all the 
freezing modes are degenerate, the coefficient of the quadratic term is given by: 

\begin{eqnarray}
T_{2} = {\cal G}^{-1}_{0,0}({\bf g}^{(1)}_{\rm d}) + 
2{\cal G}^{-1}_{0,-1}({\bf g}^{(1)}_{\rm d}) + 
{\cal G}^{-1}_{-1,-1}({\bf g}^{(1)}_{\rm d}) +
{\cal G}^{-1}_{0,0}({\bf g}^{(3)}_{\rm d}) +
2{\cal G}^{-1}_{0,1}({\bf g}^{(3)}_{\rm d}) +
{\cal G}^{-1}_{1,1}({\bf g}^{(3)}_{\rm d}).\nonumber\\
\end{eqnarray}

\noindent
We find the point in $c_{0} - (-\beta V_{e})$ plane for which $T_{2} = 
O$.\\

\section{Renormalization of the 4-Point Vertices}

\noindent
In Section 3 we locate the position at which $T_{2}$ goes through zero in the 
$c_{0} - (-\beta V_{e})$ parameter space. Our aim here is to investigate the sign 
of the coefficient $T_{4}$ of the quartic term in the effective free energy at 
this point. The first step towards investigating this question is to evaluate 
the renormalization of the four-point vertices appearing in the free energy 
expansion given in Section 2. Here we outline the calculation of the corrections 
to the four-point vertex functions.\\

\noindent
The terms quartic in the freezing mode $\xi_{\rm d}$, correspond to 
diagrams 
with four external legs labelled by the wavevectors belonging to the class 
$\{{\bf g}_{\rm d}\}$ (See Fig. 3a). Given the pair of wavevectors 
${\bf g}^{(1)}_{\rm d}$ and ${\bf g}^{(3)}_{\rm d} = -{\bf g}^{(1)}_{\rm d}$, 
one can generate the remaining two vectors in the class $\{{\bf g}_{\rm d}\}$ by 
adding to them a suitable integral multiple $m{\bf g}_{\rm f}$ with $m$ 
taking 
integral values between $-1$ and $+1$. We restrict ourselves to the lowest order 
({\it i.e.} one-loop diagrams and their ladders) fluctuation corrections coming 
from intermediate interactions between various modes. We use the renormalized 
correlation functions obtained via Dyson's equation (Sect. 3) to evaluate the 
one-loop corrections to the four-point vertex functions. We notice that for the 
four-point vertices shown in Figure 3a there are no one-loop corrections 
coming from interactions mediated by cubic vertices. This is due to the absence 
of such cubic vertices in the effective free energy which follows from the 
fact that we choose the wavevectors pertaining to the external potential in 
such a way that an integral combination of vectors in the class $\{{\bf g}_{\rm f}\}$ 
cannot be obtained from an odd combination of vectors in the class$\{{\bf g}_{\rm d}\}$. 
Thus, corrections to the bare four-point vertex functions come entirely 
from intermediate scattering processes involving the four-point vertices. 
We emphasize that the four-point vertices have two parts, $Q^{(1)}$ and 
$Q^{(2)}$ (see Eq. (4)). Clearly $Q^{(2)}$ corresponds to a more stringent 
momentum conservation condition compared to $Q^{(1)}$. Consider for instance 
a general one-loop diagram of the form shown in Figure 3b. It is easy to see 
by inspection that corrections to $Q^{(2)}$ come from such a loop only 
if ${\bf q} = {\bf g}^{(1)}_{\rm d}$. In contrast, one-loop corrections to 
$Q^{(1)}$ come from the entire range of values of ${\bf q}$.\\

\noindent
We define a function $\chi_{m_{1}m_{2}m^{\prime}_{1}m^{\prime}_{2}}$ as follows:

\begin{eqnarray*}
\chi_{m_{1}m_{2}m^{\prime}_{1}m^{\prime}_{2}} = 
\sum^{+1}_{n_{1}=-1}\sum^{+1}_{n_{2}=-1} \left[
\Pi^{(1)}_{m_{1}m_{2}n_{1}n_{2}} Q^{(1)}_{n_{1}n_{2}m^{\prime}_{1}m^{\prime}_{2}}
+ \Pi^{(2)}_{m_{1}m_{2}n_{1}n_{2}} Q^{(2)}_{n_{1}n_{2}m^{\prime}_{1}m^{\prime}_{2}}\right]
\end{eqnarray*} 
where 
\begin{eqnarray*}
\Pi^{(1)}_{l_{1}l^{\prime}_{1}k_{1}k^{\prime}_{1}} = 
\int \frac{{\rm d^2}{\bf q}}{(2\pi)^{2}} {\cal G}_{l_{1}l^{\prime}_{1}} 
({\bf q})
{\cal G}_{k_{1}k^{\prime}_{1}}(-{\bf q})
\end{eqnarray*}
and 
\begin{eqnarray*}
\Pi^{(2)}_{l_{1}l^{\prime}_{1}k_{1}k^{\prime}_{1}} =
\frac{1}{(2\pi)^{2}}{\cal G}_{l_{1}l^{\prime}_{1}}({\bf g}^{(1)}_{\rm d})
{\cal G}_{k_{1}k^{\prime}_{1}} 
(-{\bf g}^{(1)}_{\rm d}).
\end{eqnarray*}

\noindent
Summing over the ladders of one-loop diagrams (Fig. 3c), we obtain the correction 
to the vertex $Q_{m_{1}m_{2}m^{\prime}_{1}m^{\prime}_{2}}$:
\begin{eqnarray*}
\Delta V_{m_{1}m_{2}m^{\prime}_{1}m^{\prime}_{2}} = 
\sum_{n_{1},n_{2}} \tilde{Q}_{m_{1}m_{2}n_{1}n_{2}} 
\left[\delta_{m^{\prime}_{1}n_{1}} \delta_{m^{\prime}_{2}n_{2}} - 
\chi_{n_{1}n_{2}m^{\prime}_{1}m^{\prime}_{2}}\right]^{-1}
\end{eqnarray*}
where $\tilde{Q}_{m_{1}m_{2}n_{1}n_{2}} = 
Q^{(1)}_{m_{1}m_{2}l_{1}k_{1}} 
\Pi^{(1)}_{l_{1}l^{\prime}_{1}k_{1}k^{\prime}_{1}} 
Q^{(1)}_{l^{\prime}_{1}k^{\prime}_{1}n^{\prime}_{1}n^{\prime}_{2}} + 
Q^{(2)}_{m_{1}m_{2}l_{1}k_{1}}
\Pi^{(2)}_{l_{1}l^{\prime}_{1}k_{1}k^{\prime}_{1}} 
Q^{(2)}_{l^{\prime}_{1}k^{\prime}_{1}n^{\prime}_{1}n^{\prime}_{2}}$. Thus the 
corrected four-point vertex function corresponding to external legs 
labelled by $m_{1}, m_{2}, m_{1}^{\prime}$ and $m^{\prime}_{2}$ is given by 
\begin{eqnarray*}
Q^{\star}_{m_{1}m_{2}m^{\prime}_{1}m^{\prime}_{2}} = 
Q_{m_{1}m_{2}m^{\prime}_{1}m^{\prime}_{2}} +
(\Delta V)_{m_{1}m_{2}m^{\prime}_{1}m^{\prime}_{2}}
\end{eqnarray*}
with $Q_{m_{1}m_{2}m^{\prime}_{1}m^{\prime}_{2}}$ the bare four-point vertex. \\

\noindent
The relevant quantity which dictates the nature of the phase transition is, of 
course, $T_{4}$ which is the coefficient of the term quartic in $\xi_{\rm d}$ 
in the effective free energy. It is given by 

\begin{eqnarray}
T_{4} = \sum Q^{\star}_{m_{1}m_{2}m^{\prime}_{1}m^{\prime}_{2}}
\end{eqnarray}
where the summation is over those values of $m_{1}, m_{2}, m_{1}^{\prime}$ and 
$m^{\prime}_{2}$ which correspond to the vertices with external legs labelled 
by the vectors belonging to the class $\{{\bf g}_{\rm d}\}$. \\

\section{Numerical Results}

\noindent
In Figure 4, we show the renormalized $c$ and the bare 
$c_{0} (= 1/\rho_{0}c^{(2)}(q_{0}) -1)$ for different values of $-\beta V_{e}$ at which 
$T_{2} = 0$ (Eq. (15)). We notice that the values of $c$ are always larger than 
$c_{0}$. This indicates that fluctuations tend to reduce the strength of the 
correlation in the system. The enhanced effect of fluctuations with increasing 
$-\beta V_{e}$ may probably be attributed to the partial $(1-{\rm d})$ ordering 
caused by the external potential. We find that $T_{4}$ (evaluated from Eq. (16)) 
at the point $T_{2} = 0$ is positive, which is the signature of a continuous 
transition. So, the transition is continuous as found in the mean field 
calculations as well. 

\begin{center}
\includegraphics*[width=3.3in]{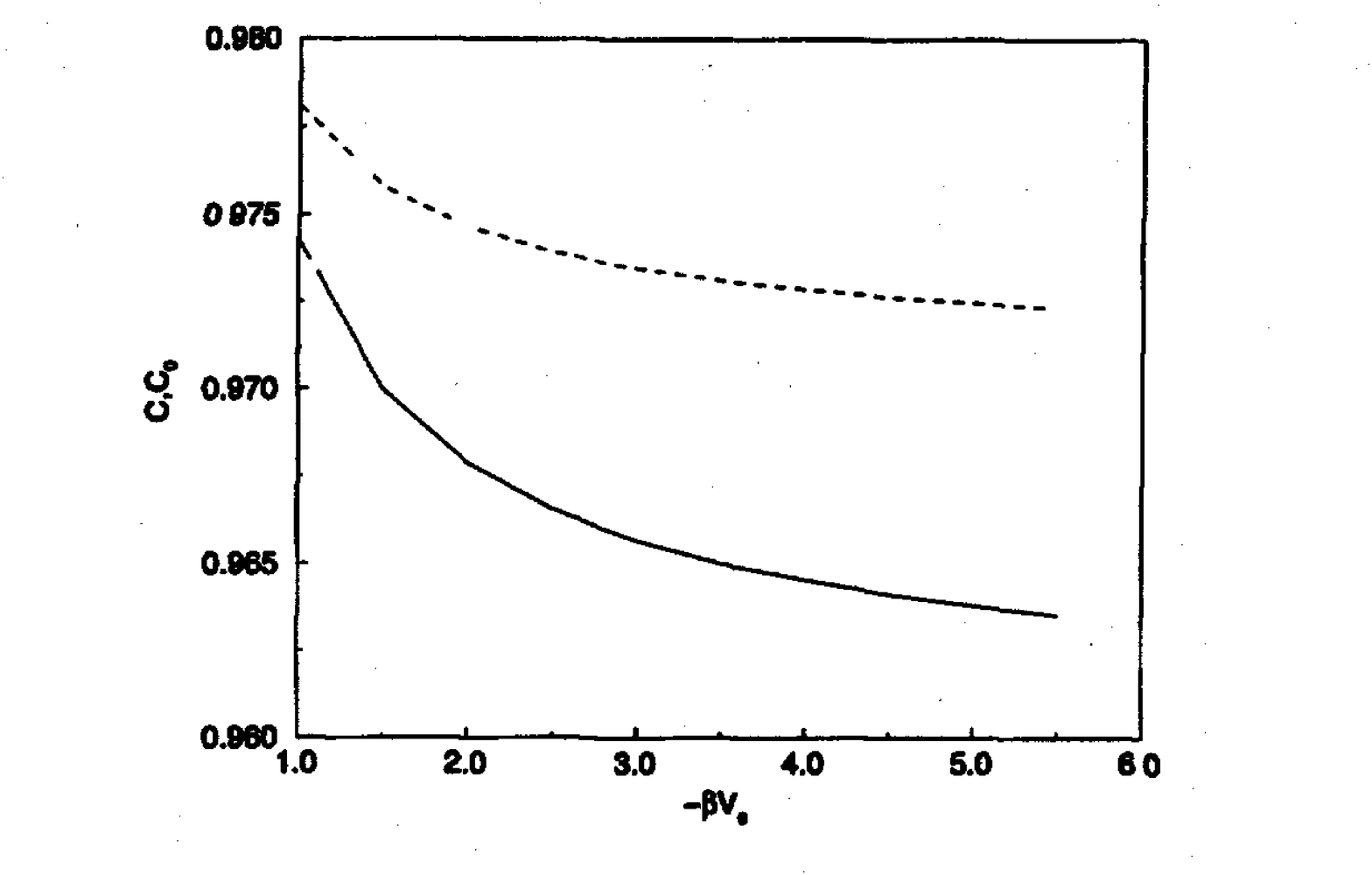}
\end{center}
\vspace*{0.3cm}
\noindent
Fig. 4. - The values of $c$ (dashed line) and $c_{0}$ (solid line) at
$T_{2} = 0$ for different $-\beta V_{e}$.

\begin{center}
\includegraphics*[width=3.2in]{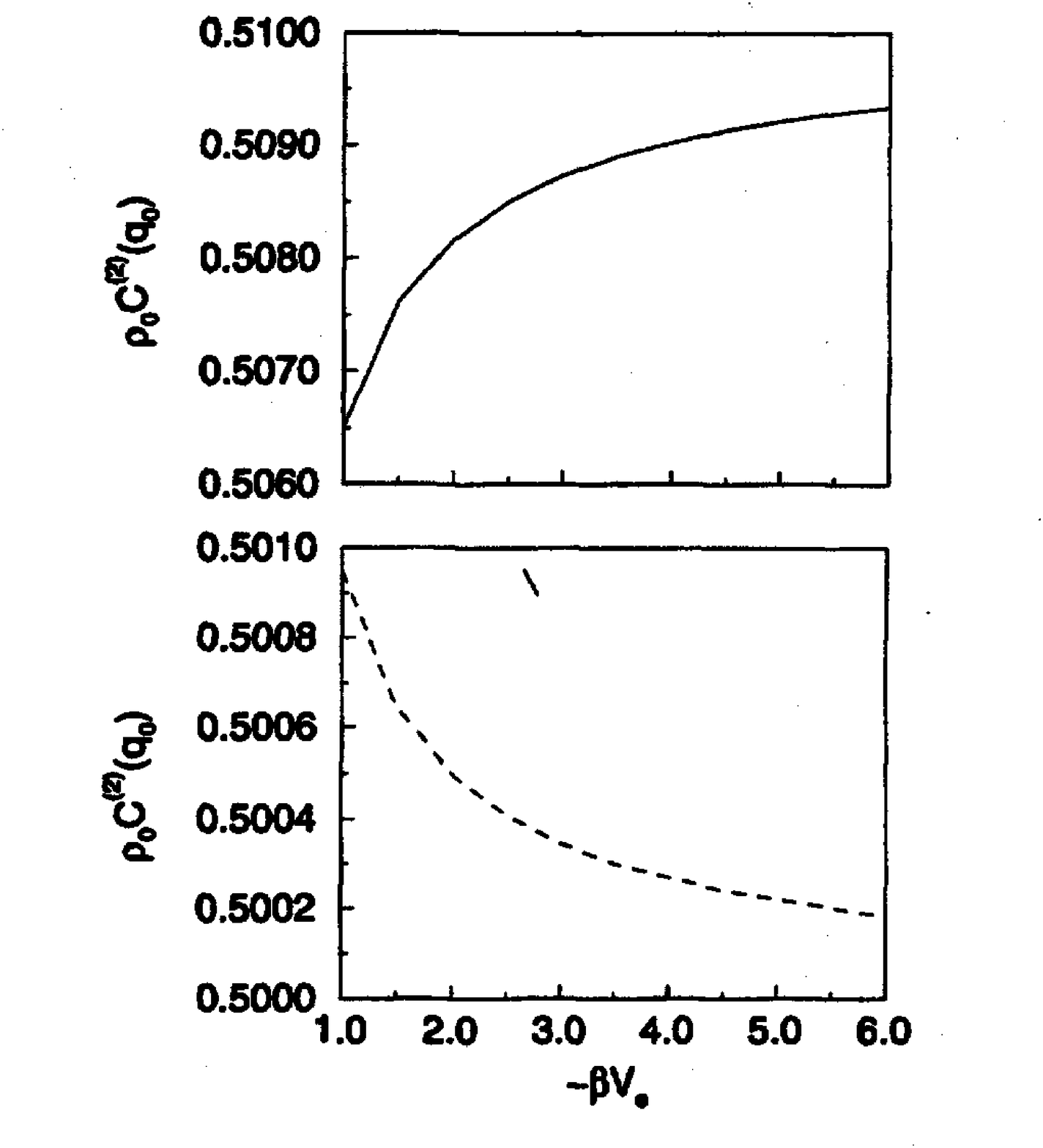}
\end{center}
\vspace*{0.2cm}
\noindent
Fig. 5. - The phase diagram, {\it i.e.}, $T_{2} = 0$ line in the 
$\rho_{0}c^{(2)} (q_{0}) - (-\beta V_{e})$ plane obtained from the 
calculations including fluctuation effects (solid line) and from the 
mean field calculations (dashed line).\\

\noindent
The phase diagram in the $\rho_{0}c^{(2)} (q_{0}) - (-\beta V_{e})$ plane is 
shown in Figure 5. Notice that as one goes to higher $-\beta V_{e}$, in 
comparison to the mean field theory the fluctuations enhance the stability of 
the liquid phase relative to the crystal phase but the transition line 
eventually saturates as in the mean field theory. However, in the mean 
field phase diagram the critical line asymptotes to 
$\rho_{0}c^{(2)} (q_{0}) = 0.5$ for large $-\beta V_{e}$ from above, 
while that in the present calculation asymptotes to 
$\rho_{0}c^{(2)} (q_{0}) = 0.509$ from below, a feature found 
in recent simulations as well [13]. Thus there is a difference in the 
curvature of the critical line in the 
$\rho_{0}c^{(2)} (q_{0}) - (-\beta V_{e})$ plane in the two cases, which 
represents a {\it qualitative difference} between the present phase 
diagram and 
the mean field phase diagram.\\

\section{Conclusion } 

\noindent 
In conclusion we have presented a theory of laser induced freezing which 
accounts approximately for the effect of fluctuations on the freezing 
transition. The freezing transition in our calculation remains continuous 
even in the presence of fluctuations for large values of the external 
modulating potential. This implies that one can perform light scattering 
experiments to look for critical opalescence indicating the presence 
of divergent static correlations of density fluctuations. \\

\noindent
The main qualitative result of this study is contained in Figure 5. 
As mentioned in Section 5, the curvature of the critical line in the 
$\rho_{0}c^{(2)}(q_{0}) - (-\beta V_{e})$ plane changes sign when 
fluctuations are taken into account. Thus, fluctuations tend to enhance 
the stability of the liquid phase relative to the crystal phase. It would 
be interesting to find out if this feature of the freezing transition is 
seen in real experiments.\\

\noindent 
Due to the inherent complexity of the problem, we have confined ourselves 
to self-consistent one-loop corrections to the self-energy and incorporated 
only the one-loop diagrams and their ladders into the renormalization of 
the four-point vertex. Furthermore, we have confined ourselves to the 
dominant corrections to the vertices coming from the low energy region 
({\it i.e}. $|{\bf q}| \sim q_{0}$) of the order parameter spectrum. At 
this level of approximation there are no significant corrections to the 
self-energy from the interactions mediated by the cubic vertices. 
However, for small values of {\bf q}, these corrections can be quite 
significant (compared to the one-loop correction mediated by the quartic 
vertices included in this paper). In future it would be certainly 
worthwhile to carry out a more systematic analysis of the role of 
fluctuations in the laser induced freezing transition.\\

\noindent
{\bf Acknowledgments}\\

\noindent
JC thanks the Council for Scientific and Industrial Research for 
financial support. We want to thank the Supercomputer Education 
and Research Centre at the Indian Institute of Science for computing 
facilities; Sriram Ramaswamy for suggesting the problem; and 
H.R. Krishnamurthy, D. Gaitonde and Pinaki Majumdar for helpful 
discussions. We thank G.I. Menon and D. Gaitonde for critical 
comments on the manuscript.

\end{document}